\documentclass[]{article}

\usepackage{widetext}
\usepackage{graphicx}
\usepackage{caption}
\usepackage{subcaption}

\usepackage{verbatim}
\usepackage{amsmath}
\usepackage{amssymb}
\usepackage{color}
\usepackage{hyperref}
\usepackage{stackrel}
\usepackage{upgreek}
\usepackage{csquotes}
\usepackage{blindtext}
\usepackage{lipsum}
\usepackage{multirow}
\usepackage{fullpage}

\definecolor{lightgray}{gray}{0.8}
\definecolor{llightgray}{gray}{0.95}

\def\##1{{\underline #1}}
\def\=#1{\underline{\underline{#1}}}
\def\+#1{\underline{\bf #1}}
\def\*#1{\breve{\bf #1}}

\def\.{\mbox{ \tiny{$^\bullet$} }}

\def\le{\left(}
\def\ri{\right)}
\def\les{\left[}
\def\ris{\right]}
\def\lec{\left\{}
\def\ric{\right\}}

\def\r#1{(\ref{#1})}

\def\eps{\varepsilon}
\def\epso{\eps_{\scriptscriptstyle 0}}
\def\etao{\eta_{\scriptscriptstyle 0}}
\def\ko{k_{\scriptscriptstyle 0}}
\def\lambdao{\lambda_{\scriptscriptstyle 0}}

\def\lambdaomin{\lambda_{\scriptscriptstyle 0, {\rm min}}}
\def\lambdaomax{\lambda_{\scriptscriptstyle 0,{\rm max}}}

\def\Eo{E_{\scriptscriptstyle 0}}

\def\00{^{(0,0)}}

\def\n{^{(n)}}

\def\ux{\hat{\#u}_x}
\def\uy{\hat{\#u}_y}

\def\Ld{L_{\rm d}}	
	
\def\Lm{L_{\rm m}}

\def\Ls{L_{\rm s}}
\def\Lt{L_{\rm t}}
\def\Lw{L_{\rm w}}
\def\Lx{L_{\rm x}}

\def\LAg{L_{\rm Ag}}
\def\LAl2O3{L_{\rm Al_{2}O_{3}}}

\def\LAu{L_{\rm Au}}
\def\LARC{L_{\rm ARC}}

\def\LMgF2{L_{\rm MgF_2}}
\def\LBSP{L_{\rm BSP}}
\def\LFSP{L_{\rm FSP}}
\def\LPd{L_{\rm Pd}}
\def\LGe{L_{\rm Ge}}

\def\epsdc{\varepsilon_{\rm dc}}

\def\sfE{{\sf E}}
\def\eg{\sfE_{\rm g}}
\def\ego{\sfE_{\rm g,min}}

\def\egmax{\sfE_{\rm g,max}}

\def\Jsc{J_{SC}}

\def\Voc{V_{oc}}

\def\sigman{\sigma_{\rm n}}
\def\sigmap{\sigma_{\rm p}}

\def\ni{n_{\rm i}}

\def\ni{n_{\rm i}}
\def\Nc{N_{\rm c}}
\def\Nv{N_{\rm v}}
\def\Nf{N_{\rm f}}
\def\NT{N_{\rm T}}

\def\sfE{{\sf E}}
\def\Ei{\sfE_{\rm i}}
\def\Ec{\sfE_{\rm c}}

\def\ET{\sfE_{\rm T}}
\def\Ev{\sfE_{\rm v}}
\def\eg{\sfE_{\rm g}}
\def\ego{\sfE_{\rm g,min}}
\def\egmax{\sfE_{\rm g,max}}

\def\egow{\sfE_{\rm g,w}}

\def\Jsc{J_{\rm sc}}

\def\Jdev{J_{\rm dev}}

\def\mun{\mu_{\rm n}}
\def\mup{\mu_{\rm p}}
\def\n0{n_{\scriptscriptstyle 0}}
\def\p0{p_{\scriptscriptstyle 0}}
\def\ni{n_{\rm i}}
\def\Nc{N_{\rm c}}
\def\Nv{N_{\rm v}}
\def\Nf{N_{\rm f}}
\def\ND{N_{\rm D}}

\def\RB{R_{\rm B}}

\def\sigman{\sigma_{\rm n}}
\def\sigmap{\sigma_{\rm p}}

\def\Voc{V_{\rm oc}}

\def\Vext{V_{\rm ext}}

\def\RB{R_{\rm B}}

\def\vthn{v_{\rm th,n}}
\def\vthp{v_{\rm th,p}}

\def\xib{\bar{\xi}}

\begin{document}

\begin{center}
 
\textbf{Optoelectronic Optimization of Graded-Bandgap Thin-Film AlGaAs
	Solar Cells }\\
	
	Faiz Ahmad,$^{1,\ast}$ Akhlesh Lakhtakia,$^{1,2}$ and Peter B. Monk$^3$\\

$^1${Pennsylvania State University, Department of Engineering Science and Mechanics, NanoMM--Nanoengineered Metamaterials Group,   University Park, PA 16802, USA}\\

$^2${Danmarks Tekniske Universitet,  Institut for Mekanisk Teknologi, Sektion for Konstruktion og Produktudvikling,
DK-2800 Kongens Lyngby, Danmark}\\

$^3${University of Delaware, Department of Mathematical Sciences,
	501 Ewing Hall,  Newark, DE 19716, USA}\\

$^\ast${Corresponding author: fua26@psu.edu}\\

\begin{abstract}

An optoelectronic optimization was carried out for an Al$_{\xi}$Ga$_{1-\xi}$As (AlGaAs) solar cell containing {(i)} an 
$n$-AlGaAs absorber layer with a graded bandgap  {and (ii)} a periodically corrugated Ag backreflector combined with  
localized ohmic Pd--Ge--Au backcontacts. The bandgap of the absorber layer was varied 
either sinusoidally or linearly.  
An efficiency of $33.1$\% with the $2000$-nm-thick $n$-AlGaAs absorber layer is predicted with linearly graded bandgap along
with silver backreflector and localized ohmic backcontacts, in comparison to $27.4$\% efficiency obtained 
with homogeneous bandgap  and a continuous ohmic backcontact. Sinusoidal grading of the
bandgap {is predicted to enhance} the maximum efficiency
to $34.5\%$.
Thus,  grading the bandgap of the absorber layer, along with a periodically corrugated
Ag backreflector and localized ohmic Pd--Ge--Au backcontacts can help realize ultrathin and high-efficient 
AlGaAs solar cells for terrestrial applications.  

\end{abstract}

\end{center}
 
	\def\doubleunderline#1{\underline{\underline{#1}}}
	\renewcommand\vec{\mathbf}

\section{Introduction}

Highly efficient and cost-effective solar cells made ecoresponsibly~\cite{Vezzoli}
of   {Earth-abundant} materials with low
after-use disposal environmental cost 
are necessary for sustainability~\cite{Hawken}. With crystalline-silicon (Si) delivering about $26\%$ efficiency 
and multicrystalline-Si about $22\%$ efficiency, Si is the photovoltaic material of choice
for solar-photovoltaic modules deployed in solar parks and on rooftops \cite{Green2018,Green-PE2019}. 
With somewhat higher efficiency and significantly lower weight-to-power ratio \cite{Geelen1997}, 
gallium arsenide (GaAs) is the current market leader for solar cells deployed for extra-terrestrial applications, 
but it is prohibitively expensive for terrestrial applications~\cite{Horowitz}.

There are two options to reduce the cost of the GaAs solar cell.
The first option is the reduction of the thickness of the GaAs absorber  layer~\cite{Kayes2011,Lee2015, Vandamme2015}. 
Not only will that option reduce material usage, but it will also enhance
manufacturing throughput. 
However, a thinner absorber layer will reduce the absorption of incident solar photons. 
Back-surface modifications such as plasmonic nanostructures~\cite{XIAO2018, Lee2015, Wang2014}, localized ohmic backcontacts~\cite{Vandamme2015}, and highly reflective backreflectors~\cite{Vandamme2015} have been investigated 
to tackle the problem of low absorption in ultrathin GaAs solar cells, but enhanced photon trapping does not 
necessarily translate into higher efficiency \cite{Ahmad-SPIE2018,Ahmad2019}. 

The second option is to grade the bandgap in the absorber layer by adding aluminum (Al)
and controlling the compositional ratio of Al to gallium (Ga)~\cite{Hutchby1975,Dharmadasa2005}. Bandgap grading of 
the resulting Al$_{\xi}$Ga$_{1-\xi}$As (AlGaAs) absorber layer will allow photon absorption over a wider frequency 
range. Also, bandgap grading  will increase efficiency by
creating a drift electric field that will accelerate photogenerated
holes towards the $p$-$n$ junction   {in the solar cell}~\cite{Hutchby1975}. Linear   {bandgap grading} has been shown experimentally 
to increase the open-circuit voltage $\Voc$ in AlGaAs solar cells~\cite{Dharmadasa2015}, which should assist 
in enhancing the efficiency $\eta$; however, suboptimal bandgap grading can reduce the short-circuit current 
density $\Jsc$ to offset the increase in $\Voc$.

A recent theoretical study on CIGS solar cells shows that $\Voc$ can be enhanced while maintaining or even 
enhancing  $\Jsc$~\cite{Ahmad2019}, by  optimally grading the bandgap of the absorber layer.
Motivated by these results, we combined both  options, i.e.,
thinning~\cite{Kayes2011,Lee2015, Vandamme2015} and bandgap grading
~\cite{Hutchby1975,Dharmadasa2005,Dharmadasa2015}
of the absorber layer  in a coupled optoelectronic model~\cite{Anderson-JCP}
of a thin-film AlGaAs solar cell. We then used the model to determine
optimal geometric and bandgap-grading parameters to maximize $\eta$.

The thickness of the AlGaAs absorber layer was allowed to vary from
100~nm to 2000~nm, and the bandgap was allowed
to vary either linearly or sinusoidally along the thickness direction.
In addition, we incorporated 
a highly reflective periodically corrugated 
silver (Ag) backreflector and \textit{localized} ohmic 
backcontacts of palladium (Pd), germanium (Ge), and gold (Au) trilayers~\cite{Vandamme2015}, with the  
areal ratio $\zeta\in[0,1]$ of     Pd--Ge--Au and  Ag being a geometric parameter for optimization. 
When $\zeta=1$, the Ag backreflector is  absent  while a Pd--Ge--Au
trilayer extends across the entire back surface as is typical for 
a GaAs solar cell~\cite{Bauhuis2009,Kayes2011}.

The coupled optoelectronic model has an optical part and an electrical part. In the optical part, 
the rigorous coupled-wave approach (RCWA)~\cite{GG,ESW2013} is used to determine the   {electron--hole-pair} 
generation rate in the semiconductor layers of the solar cell~\cite{Anderson-JCP,Ahmad2019}, assuming 
normal illuminationby unpolarized polychromatic light endowed with the AM1.5G solar spectrum~\cite{SSAM15G}. 
In the electrical part, the   {electron--hole-pair} generation rate is used as an input to the one-dimensional (1D) 
drift-diffusion equations~\cite{Fonash, Jenny_Book} applied to the semiconductor layers. These equations
are solved using a hybridizable discontinuous Galerkin (HDG) scheme~\cite{Lehrenfeld, CockburnHDG,FuQiuHDG, 
Brinkman, Brezzi2002} to determine the  current density $\Jdev$ and the electrical power density $P=\Jdev\Vext$ as 
functions of the bias voltage $\Vext$ under steady-state
conditions. In turn, the   {$\Jdev$-$\Vext$} and the   {$P$-$\Vext$} curves yield $\Jsc$, $\Voc$. and $\eta$. 
Finally, the differential evolution algorithm (DEA)~\cite{DEA} is used to maximize $\eta$
as a function of various geometric and bandgap-grading parameters. 

The structure of this paper is as follows.  Section~\ref{Sec:optoelec-model} contains the optical and 
the electrical descriptions of the AlGaAs solar cell.  
As implementation details for the optical~\cite{Ahmad-SPIE2018,Anderson-JCP} and the 
electronic parts \cite{Anderson-JCP,Ahmad2019} of the model as well as the DEA \cite{Solano,Solano-err}
for solar-cell problems have been published, we have not provided them in this paper.
Section~\ref{Sec:OptoElecRes} divided into four subsections. 
The efficiency of the solar cell with a 2000-nm-thick GaAs 
layer as predicted by the model is compared with experimental results~\cite{Bauhuis2009}
in Sec.~3.\ref{model-valid}. 
The effects of the periodically corrugated Ag backreflector along with 
localized ohmic Pd--Ge--Au  backcontacts on the performance of the GaAs solar
cells are discussed in Sec.~3.\ref{Sec:localized-contact}. Next,
optimal results for solar cells with a homogeneous AlGaAs absorber layer (Sec~3.\ref{Sec:homo-LBC}),
an AlGaAs absorber layer with linearly graded bandgap (Sec.~3.\ref{Sec:Linearly-graded}),
and an AlGaAs absorber layer with sinusoidally graded bandgap (Sec.~3.\ref{Sec:Sinusoidally-graded}) are provided, 
each solar cell possessing a periodically corrugated Ag backreflector along with 
localized ohmic Pd--Ge--Au  backcontacts. 
The paper concludes with some remarks in Sec.~\ref{sec:conc}.

\section{Optical and Electrical Descriptions} \label{Sec:optoelec-model}

\begin{figure}[h]
	\centering
		\includegraphics[width=1.2\columnwidth]{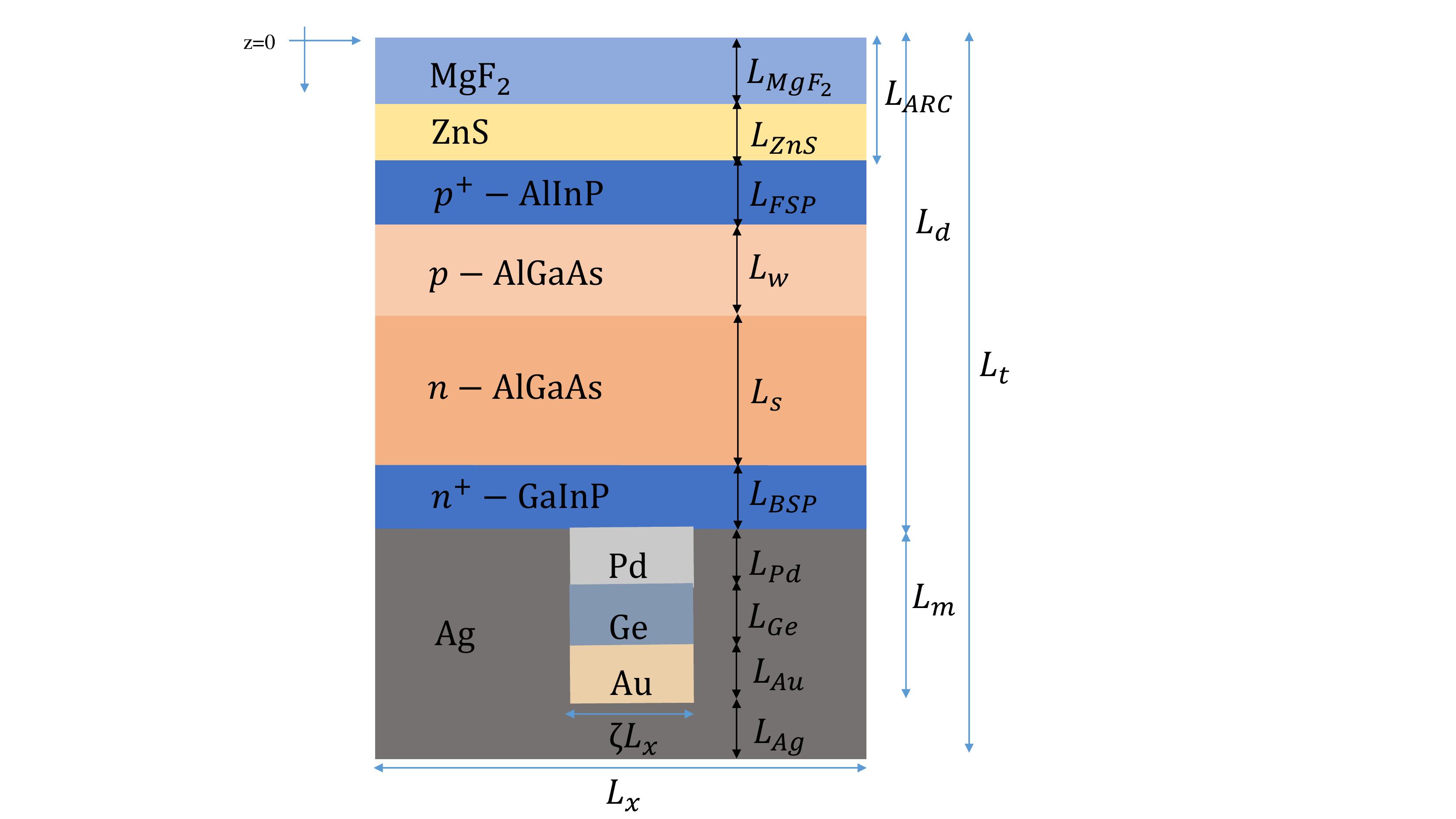}    
	\caption{Schematic of the reference unit cell $\cal R$ of the AlGaAs solar cell.  
		\label{figure1}}
\end{figure}

The solar cell occupies the region
${\cal X}:\left\{(x,y,z) \vert -\infty\right.$ $\left. <x<\infty, -\infty<y<\infty, 0<z<\Lt\right\}$, with the
half spaces $z<0$ and $z>\Lt$ occupied by air. The reference unit cell, identified as
${\cal R}:\left\{(x,y,z) \vert -\Lx/2<x<\Lx/2,  -\infty<y<\infty, 0<z<\Lt\right\}$, is schematically 
depicted in Fig.~\ref{figure1}.

The region $0<z<\LMgF2=110$~nm is occupied by magnesium fluoride (MgF$_2$)~\cite{mgf2}
and the region $\LMgF2<z<\LARC=150$~nm by zinc sulfide (ZnS)~\cite{ZnS}, the two layers 
collectively functioning to reduce light reflection~\cite{Bauhuis2009}. 
The region $\LARC<z<\LARC+\LFSP$ is a 20-nm-thick  front-surface passivation (FSP) layer of
$p^{+}$-Al$_{0.51}$In$_{0.49}$P (hereafter referred as AlInP)~\cite{AlInP}  to reduce the 
front-surface 
recombination rate and thereby improve $\Jsc$~\cite{NREL-Report-1999}.
Next, homogeneous $p$-Al$_{\xib}$Ga$_{1-\xib}$As~\cite{Aspnes1986} with
fixed $\xib$ occupies the 50-nm-thick region
$\LARC+\LFSP<z<\LARC+\LFSP+\Lw$ to form a $p$-$n$ junction with an
$n$-Al$_{\xi}$Ga$_{1-\xi}$As~\cite{Aspnes1986} absorber layer  of thickness $\Ls\in[100,2200]$~nm. 
The quantity $\xi$ is taken to be dependent on $z$
in this paper. With $\Ld=\LARC+\LFSP+\Lw+\Ls+\LBSP$,
the region $\Ld-\LBSP<z<\Ld$ of thickness $\LBSP=20$~nm  is a back-surface passivation (BSP) layer
of $n^+$-Ga$_{0.49}$In$_{0.51}$P (hereafter referred as GaInP)~\cite{GaInP} to reduce the back-surface 
recombination rate and thereby improve $\Jsc$~\cite{NREL-Report-1999, Roos-1978}.

The region $\Ld < z < \Ld+\Lm$ in ${\cal R}$ has a complicated morphology.
A Pd--Ge--Au triple layer of width $\zeta\Lx$, $\zeta\in(0,1)$, along the $x$ axis
serves as the localized ohmic back contact~\cite{Vandamme2015} comprising a Pd layer
of thickness $\LPd=20$~nm~\cite{JohnsonChristy}, a Ge layer of thickness $\LGe=50$~nm~\cite{Jellison1992}, 
and an Au layer of thickness $\LAu=100$~nm~\cite{JohnsonChristy}. The remainder of the region $\Ld < z < \Ld+\Lm$
is occupied by Ag~\cite{JohnsonChristy} for  optical reflection.
Finally, the region $\Ld+\Lm < z<\Ld+\Lm+\LAg=\Lt$,
$\LAg=100$~nm, is occupied by Ag  serving as an optical backreflector.

The linear variation of bandgap in the $n$-AlGaAs absorber layer  was modeled as 
  {\cite{Ahmad-SPIE2018,Ahmad2019}}
\begin{eqnarray} 
\nonumber
&&\eg(z)=\ego 
\\ &&+A\left(\egmax-\ego\right)\frac{z-\left(\LARC+\LFSP+\Lw\right)}{\Ls}\, ,
\nonumber
\\
&&
z\in\left[\LARC+ \LFSP+\Lw, \LARC+\LFSP+\Lw+\Ls\right]\, ,
\label{Eqn:Linear-bandgap1}
\end{eqnarray}
where $\ego$ is the minimum
bandgap, $\egmax$ is the maximum bandgap,
and $A$ is an amplitude (with $A = 0$ representing a homogeneous AlGaAs layer). 
The bandgap is thus minimum at the front face $z=\LARC+\LFSP+\Lw$
and maximum at the back face  $z=\LARC+\LFSP+\Lw+\Ls$ of the absorber layer.
The reverse grading (i.e., maximum at   $z=\LARC+\LFSP+\Lw$
and minimum at    $z=\LARC+\LFSP+\Lw+\Ls$) did not give satisfactory results.

The sinusoidal variation of the bandgap in the $n$-AlGaAs absorber layer  
was modeled as   {\cite{Ahmad-SPIE2018,Ahmad2019,Anderson2018}}
\begin{align} 
\nonumber
&\eg(z)=\ego +A\le \egmax-\ego\ri \, 
\\
\nonumber
&\times
\lec \frac{1}{2}\, \les \sin\le 2\pi K \frac{z-\left(\LARC+\LFSP+\Lw\right)}{\Ls}-2\pi \psi\ri +1\ris\, \ric^{\alpha} \, , 
\\[5pt]
&\qquad  z\in\left[\LARC+\LFSP+\Lw, \LARC+\LFSP+\Lw+\Ls\right]\, ,
\label{Eqn:Sin-bandgap}
\end{align}
where $\psi\in [0, 1]$ describes a relative phase shift, $K$ is the number of periods in the AlGaAs layer, 
and $\alpha> 0 $ is a
shaping parameter.  
The parameter $\xib$  for the homogeneous $p$-AlGaAs  layer governs the bandgap
$\egow$ in that layer.  Optical spectra of the relative permittivities of  all materials used in the solar cell
are provided in Appendix \ref{Appendix A}.

The RCWA~\cite{GG, ESW2013} was used the calculate the electric field phasor ${\#E}(x,z,\lambdao)$ everywhere inside 
the solar cell as a result of illumination by a monochromatic plane wave normally incident on the plane $z=0$ from 
the half space $z<0$,  $\lambdao$ being the free-space wavelength.
   The electric field phasor of the incident plane wave was taken as
${\#E}_{\rm inc}(z,\lambdao)=
{\Eo} \frac{\ux+\uy}{\sqrt{2}}\exp\left(i\ko{z}\right)\,$
with $\Eo=4\sqrt{15 \pi}$~V~m$^{-1}$.
With the assumption that every absorbed photon excites an electron-hole pair,
the $x$-averaged   {electron--hole-pair} generation rate was calculated as~\cite{Ahmad2019} 	%
\begin{align} 
\nonumber
&G(z)=\frac{\etao}{\hbar\Eo^2 }\frac{1}{\Lx}  \int_{-\Lx/2}^{\Lx/2}\Bigg[ \int_{\lambdaomin}^{\lambdaomax} {\rm Im}\{\eps(x, z, \lambdao)\}
\left\vert\#E(x, z, \lambdao)\right\vert^2\,
\\[8pt]
&\qquad\qquad\qquad\times 
S(\lambdao)  \,  d\lambdao\Bigg] dx  \,
\label{G1D-def}
\end{align}
for $z\in\les \LARC, \Ld \ris$, where  $\hbar$ is  the reduced Planck constant, $\etao= 120\pi~\Omega$ is the intrinsic impedance
of free space, $S(\lambdao)$ is the AM1.5G solar spectrum~\cite{SSAM15G},
$\lambdaomin = 300$~nm, and    {$\lambdaomax = \left(1240/\ego\right)$~nm with
$\ego$ in eV.}
For use in the  electrical part of the model, $G (z)$ contains the effects of highly reflective
Ag, the localized ohmic Pd--Ge--Au backcontacts, and  the MgF$_2$/ZnS double-layer antireflection coating. 
The $x$-averaging is justified since  the charge carriers  generally flow along the $z$ axis because the 
solar cell operates under the influence of a bias voltage $\Vext$
applied along the same axis; furthermore, $\Lx \sim$500~nm is minuscule in comparison to the lateral 
dimensions of the solar cell.

The region $\LARC<z<\Ld$ contains four semiconductor layers: the $p^{+}$-AlInP FSP layer, the $p$-AlGaAs  layer, the
$n$-AlGaAs absorber layer, and the $n^{+}$-GaInP BSP layer. All four were incorporated
in the electrical part~\cite{Ahmad2019, Anderson-JCP} of the model, as all four contribute to charge-carrier generation. 
Electrical parameters used for all four semiconductor   {layers~\cite{Vurgaftman2001, Gudovskikh2007, Adachi1985, Adachi-book, ioffe}} are provided in Appendix~\ref{Appendix A}.

As the focus here is not on how the solar cell interfaces with an external 
circuit,   both terminals were considered to be ideal ohmic contacts. We used a 1D drift-diffusion 
model~\cite{Jenny_Book, Fonash, Brezzi2002} to investigate the transport of electrons and holes.  
The bandgap-dependent electron affinity $\chi(z)$, the conduction band density of states $\Nc(z)$, the
valence band density of states $\Nv(z)$, the electron mobility $\mun(z)$, the hole mobility $\mup(z)$, 
and the DC relative permittivity $\epsdc(z)$ were incorporated in the electrical calculations.
Also, we incorporated all three recombination processes:   {radiative, Shockley--Read--Hall, 
and} Auger~\cite{Jenny_Book, Fonash}. The Shockley--Read--Hall recombination rate was taken to depend
on $\eg(\xi)$ as the trap/defect density $\Nf$, the electron  thermal speed
$\vthn$, and the hole thermal speed $\vthp$ are $\xi$ dependent in both Al$_\xi$Ga$_{1-\xi}$As layers. However, the
radiative recombination and Auger recombination rates were considered independent of the bandgap due to 
lack of available data. The electrical part yields values of $\Jsc$, $\Voc$,   
the fill factor $FF$~\cite{Jenny_Book}, and $\eta$.

The DEA \cite{DEA} was used to maximize $\eta$ with respect to certain geometric and bandgap parameters, 
using a custom algorithm implemented with MATLAB\textsuperscript{\textregistered} version R2019a.

\section{Numerical results and discussion}\label{Sec:OptoElecRes}

\subsection{GaAs solar cell}\label{Sec:Ref-conventional-cell}

\subsubsection{Model validation}\label{model-valid}
First, we validated our coupled optoelectronic model by comparison with the experimental results for the MgF$_2$/ZnS/AlInP/$p$-GaAs/$n$-GaAs/GaInP/Pd--Ge--Au solar cell containing a $\Ls=2000$-nm-thick homogeneous
GaAs layer~\cite{Bauhuis2009}; i.e., without Ag 
($\LAg=0$ and $\zeta=1$ in reference to Fig.~\ref{figure1}),   $\xib= 0$,   {and}
$\xi(z)\equiv 0\,\forall
z\in[\LARC+\LFSP+\Lw,\LARC+\LFSP+\Lw+\Ls]$.

Values of  $\Jsc$, $\Voc$,   $FF$, and $\eta$ obtained from our model are provided in Table~\ref{tab--ref-results}, 
as also are the corresponding experimental data~\cite{Bauhuis2009}.  The model predictions are in reasonable 
agreement with the experimental data. Furthermore, the model-predicted efficiency of $27.4\%$
is close to the highest efficiency ($27.6\%$) reported \cite{Kayes2011} for GaAs solar cells, but geometric data is not available in Ref.~\cite{Kayes2011} for a proper comparison with the model predictions.

Parenthetically, no interface defects were taken into consideration in the model, suggesting that 
all the experimentally observed characteristics can be accounted   {\cite{Gudovskikh2007}} for
by the bulk properties of MgF$_2$, ZnS, AlInP, $p$-GaAs, $n$-GaAs, GaInP, Pd, Ge, and Au.

\begin {table}[h]
\caption {\label{tab--ref-results}
	{Comparison of   $\Jsc$, $\Voc$,   FF, and $\eta$ predicted by
		the coupled optoelectronic model 
		for a GaAs solar cell with a homogeneous GaAs absorber layer (i.e., $A=0$)
		with  experimental data \cite{Bauhuis2009} for $\LAg=0$, $\zeta=1$,  $\Ls=2000$~nm, and $\xi=0$ . 
		}
}
\begin{center}
	\begin{tabular}{|p{2.2cm}|p{.8cm}|p{0.6cm}|p{0.6cm}|p{0.6cm}|}
		\hline \hline
  & $\Jsc$ &$\Voc$ & $FF$&$\eta$  \\
	&(mA&(V)&(\%)& (\%)\\
 &cm$^{-2})$&&& \\
		\hline 
Model     &29.8&1.081& 85.1& 27.4\\ 	\hline
Experiment \cite{Bauhuis2009} &29.5&1.045&84.6&26.1	\\ 
  \hline
		\hline 
	\end{tabular}
\end{center}

\end {table}	

\subsubsection{Effect of \textit{localized} ohmic backcontacts}\label{Sec:localized-contact}
Typically, the Ag backreflector is  absent  while a Pd--Ge--Au
trilayer extends across the entire back surface  of a  GaAs solar cell, i.e., $\zeta=1$
\cite{Bauhuis2009,Kayes2011}. Therefore, next
we considered the effect of the localization of ohmic backcontacts by including   Ag for better 
optical backreflection~\cite{Vandamme2015}. We maximized  $\eta$ as a function
of $\Lx\in[100,1000]$~nm and $\zeta\in\les0.05,1\ris$, for $\Ls\in\lec100,1000,2000\ric$~nm,
$\xib=0$, and $\xi(z)\equiv 0\,\forall
z\in[\LARC+\LFSP+\Lw,\LARC+\LFSP+\Lw+\Ls]$.

For all three values of $\Ls$, the efficiency was found to be maximum
for $\Lx=510$~nm and $\zeta=0.05$.
Values of  $\Jsc$, $\Voc$, $FF$, and $\eta$ obtained from the coupled optoelectronic model
are provided in Table~\ref{tab-LocOhmCont}. The effect of the inclusion of the Ag backreflector 
to localize the ohmic  Pd--Ge--Au backcontacts is to increase $\Jsc$. However, that increase is
more for smaller $\Ls$. At the same time, $\Voc$ decreases significantly for $\Ls=100$~nm,
but   {it} does not change for the two higher values of $\Ls$. As a result, the efficiency is enhanced from 
$16.5\%$ to $18.3\%$ (a relative enhancement of $10.9\%$) for $\Ls=100$~nm, but
from $27.4\%$ to just $28.0\%$ (a relative enhancement of $2.1\%$)
for $\Ls=2000$~nm. In other words, the
effect of localized ohmic backcontacts on $\eta$ 
is significant for thin absorber layers but less pronounced for thick absorber layers. 

The value of $\zeta=0.05$ is in accord with the experimental and theoretical findings of
Vandamme~\textit{et al.}~\cite{Vandamme2015}.  
Hence, we ensured that $\zeta\geq0.05$  for optimization of AlGaAs solar cells.

\begin {table}[h]
\caption {\label{tab-LocOhmCont}
	{$\Jsc$, $\Voc$,   FF, and $\eta$ predicted by the coupled optoelectronic model   
	for a GaAs solar cell with a homogeneous GaAs absorber layer (i.e., $A=0$), when 
	Ag is either absent ($\zeta=1$) or not ($\zeta<1$) and
$\Ls\in\{200, 1000, 2000\}$~nm.}
}
\begin{center}
	\begin{tabular}{|p{0.8cm}|p{0.6cm}|p{0.6cm}|p{0.6cm}|p{.8cm}|p{0.8cm}|p{0.8cm}|p{0.8cm}|}
		\hline \hline
	$\Ls$ &$\Lx$&$\zeta$&$\LAg$& $\Jsc$ &$\Voc$ & $FF$&$\eta$  \\
		(nm)&(nm)&&(nm)&(mA&(V)&(\%)& (\%)\\
		&&&&cm$^{-2})$&&& \\
		\hline
	100 & - &1 &0& 17.3 &1.132 & 84.2&16.5\\ \hline
	100	 &510&0.05&100&19.7&1.093&85.1&18.3	\\ \hline 

	1000	 &-&1&0&28.3&1.089& 85.1& 26.3\\ 	\hline
	1000	 &510&0.05&100&29.4&1.090&85.0&27.3	\\ \hline
	
	2000	 &-&1&0&29.8&1.081& 85.1& 27.4\\ 	\hline
	2000	 &510&0.05&100&30.4&1.081&85.1&28.0	\\ 
		\hline \hline
	\end{tabular}
\end{center}
\end {table}	

\subsection{Optimal AlGaAs solar cell: Homogeneous bandgap}\label{Sec:homo-LBC}

Next, we considered the optoelectronic optimization of the solar cell with a
homogeneous $n$-AlGaAs absorber layer (i.e., $A=0$), a periodically corrugated Ag 
backreflector, and localized ohmic Pd--Ge--Au backcontacts.  Whereas $\LAg=100$~nm
was fixed, the parameter space for optimizing $\eta$ was chosen
	as: $\egow\in[1.424,2.09]$~eV, $\ego\in[1.424,2.09]$~eV, 
$\Lx\in[100,1000]$~nm,  
and $\zeta\in[0.05,1]$. The common allowed range
 of $\egow$ and $\ego$ is consistent with $\xi\in[0,0.8]$.
Optimization was done for several discrete
values of $\Ls$ ranging from $100$~nm to $2000$~nm.

Values of  $\Jsc$, $\Voc$, $FF$, and $\eta$ predicted by the coupled optoelectronic   
model are presented in Table~\ref{tab-homo-LBC} for seven different values of $\Ls$. The values of $\egow$, 
$\ego$,   $\Lx$, and $\zeta$  for the optimal designs are also provided in the same table.

For the thinnest $n$-AlGaAs absorber layer ($\Ls=100$~nm), the maximum efficiency predicted is $18.5\%$ 
with $\egow=2.09$~eV ($\xib=0.8$), $\ego=1.424$~eV ($\xi=0$), and $\Lx=500$~nm. The values of $\Jsc$, $\Voc$, and $FF$ 
corresponding to this optimal design are   {$18.9$~mA~cm$^{-2}$}, 
  {$1.149$}~V, and   {$85.1\%$}, respectively. 
For this design, the $n$-AlGaAs   layer is really a $n$-GaAs layer but the
the $p$-AlGaAs layer is different from a $p$-GaAs layer.
If $\xib=0$ were to be fixed (i.e., the $p$-AlGaAs layer were to be
replaced by a $p$-GaAs layer,  the efficiency would be slightly less at $\sim18.3\%$ (Table~\ref{tab-LocOhmCont}).

For the thickest $n$-AlGaAs absorber layer ($\Ls=2000$~nm), the maximum 
  {efficiency} predicted is $28.8\%$ 
with $\egow=2.09$~eV ($\xib=0.8$), $\ego=1.424$~eV, and $\Lx=500$~nm. The values of $\Jsc$, $\Voc$, and $FF$ 
corresponding to this optimal design are   {$30.2$~mA~cm$^{-2}$}, $1.090$~V, and $87.3\%$, respectively.
Again, for this optimal design, the $n$-AlGaAs  layer is really a $n$-GaAs layer but the $p$-AlGaAs   
layer is different from a $p$-GaAs layer. If the the $p$-AlGaAs layer were to be
replaced by a $p$-GaAs layer, the efficiency would decrease somewhat to $\sim$28\% (Table~\ref{tab-LocOhmCont}).

Regardless of the value of $\Ls$, the optimal design in Table~\ref{tab-homo-LBC} has $\Lx=505\pm5$~nm and
$\zeta=0.05$, similar to the optimal design for its GaAs counterpart
(Table~\ref{tab-LocOhmCont}).  Even lower values of $\zeta$ would give higher efficiencies but the
localized ohmic Pd--Ge--Au backcontacts are necessary because of superior electron-collection 
capability~\cite{Vandamme2015}. Also, both $\ego$ and $\egow$ are independent of $\Ls$
in  Table~\ref{tab-homo-LBC}, $\ego$ being at its minimum allowed
value and $\egow$ at its maximum allowed value.

\begin {table}[h]
\caption {\label{tab-homo-LBC} 
	Predicted parameters of the optimal AlGaAs solar cell with a specified value of $\Ls\in[100,2000]$~nm, 
	when the $n$-AlGaAs absorber layer is homogeneous ($A=0$), $\LAg=100$~nm, and $\zeta<1$. }
\begin{center}
	\begin{tabular}{ |p{0.5cm}| p{0.5cm}|p{0.6cm}|p{0.5cm}|p{0.5cm}|p{0.7cm}|p{0.6cm}|p{0.5cm}|p{0.5cm}|}
		\hline
		\hline
		$\Ls$ & $\egow$&$\ego$ & $\Lx$ &$\zeta$ &
		$\Jsc$ & $\Voc$&$FF$&$\eta$  \\
		(nm) & (eV)& (eV)& (nm)& & (mA & (V)& (\%)& (\%) \\
		&	& & &    &cm$^{-2}$) & & \\
		\hline
		100 & 2.09 &1.424 &500 &0.05 &18.9&1.149&85.1&18.5	\\  \hline
		200 & 2.09 &1.424 &510 &0.05 & 21.6 &1.128 & 85.2&20.7\\ \hline
		300 & 2.09 &1.424 &502 &0.05 & 24.1 &1.124 & 85.7&23.2\\ \hline
		400 & 2.09 &1.424 &510 &0.05 & 25.8 &1.119 & 86.5&24.9\\   \hline 
		500 & 2.09 &1.424 &500 &0.05 & 27.0 &1.117 & 86.3&26.1\\   \hline
		1000 &2.09 &1.424 &500 &0.05 & 29.2 &1.104 & 87.0&28.1\\   \hline
		2000 &2.09 &1.424 &510 &0.05 & 30.2 &1.090 & 87.3&28.8\\  			
		\hline
		\hline

	\end{tabular}
\end{center}
\end {table}
	
\subsection{Optimal AlGaAs solar cell: Linearly graded bandgap} \label{Sec:Linearly-graded}

\subsubsection{Optimal designs}\label{od-linear}
Next, we considered the maximization of $\eta$ when the bandgap of the $n$-AlGaAs 
absorber layer is linearly  graded according to Eq.~(\ref{Eqn:Linear-bandgap1}), $\LAg=100$~nm,
and $\zeta\ne 1$. 
The parameter space used for optimizing $\eta$ was chosen as:  
$\egow\in[1.424, 2.09]$~eV, $\ego\in[1.424, 2.09]$~eV, $\egmax\in[1.424, 2.09]$~eV, $A\in[0,1]$,
  {$\Lx\in[100, 1000]$~nm},  and $\zeta\in[0.05,1]$. The common allowed range
 of $\egow$, $\ego$, and $\egmax$ is consistent with $\xi\in[0,0.8]$.

Values of  $\Jsc$, $\Voc$, $FF$, and $\eta$ for the optimal designs are
 presented in Table~\ref{tab-linear-grading} for seven different values of $\Ls$.
The corresponding values of $\egow$, $\ego$, $\egmax$, $A$, $\Lx$, and $\zeta$  are also provided in the same table.

For the thinnest $n$-AlGaAs absorber layer ($\Ls=100$~nm), the maximum efficiency predicted is $21.0\%$ 
with $\egow=1.424$~eV ($\xib=0$), $\ego=1.424$~eV ($\xi=0$), $\egmax=1.98$~eV ($\xi=0.45$), $A=0.99$, and $\Lx=500$~nm. 
A relative enhancement of $13.5$\% over the maximum efficiency $18.5\%$ in Table~\ref{tab-homo-LBC} 
for the homogeneous absorber layer of the same thickness is predicted.  
The values of $\Jsc$, $\Voc$, and $FF$ corresponding to the optimal design are 
  {$16.8$~mA~cm$^{-2}$}, $1.399$~V, 
and $89.3\%$, respectively.

For the thickest $n$-AlGaAs absorber layer ($\Ls=2000$~nm), the maximum efficiency predicted is $33.1\%$ 
with $\egow=1.424$~eV ($\xib=0$), $\ego=1.424$~eV ($\xi=0$), $\egmax=1.98$~eV ($\xi=0.45$),   {$A=1$}, and $\Lx=500$~nm.  
The values of $\Jsc$, $\Voc$, and $FF$ 
corresponding to this optimal design are   {$24.7$~mA~cm$^{-2}$}, $1.507$~V, and $88.8\%$, respectively.
A relative enhancement of $14.9$\% is predicted with linear bandgap grading of the $n$-AlGaAs absorber layer over the optimal 
efficiency of $28.8\%$ with the homogeneous $n$-AlGaAs absorber layer in Table~\ref{tab-homo-LBC}. For this optimal design, the $p$-AlGaAs  
layer is really a $p$-GaAs layer, but the $n$-AlGaAs absorber layer is different from a $n$-GaAs absorber layer.
Although $\Voc$ is significantly higher with the linearly graded bandgap compared to the homogeneous bandgap  (Table~\ref{tab-homo-LBC}),   $\Jsc$ is
lower with the linearly graded bandgap.

Similar to the data for the homogeneous absorber layer provided in Table~\ref{tab-homo-LBC}, 
the optimal designs in Table~\ref{tab-linear-grading} 
have $\Lx=505\pm5$~nm and $\zeta=0.05$, regardless of the value
of $\Ls$. Also, both $\ego$ and $\egow$ are independent of $\Ls$
in Table~\ref{tab-linear-grading}, just as in  Table~\ref{tab-homo-LBC}. For both homogeneous
and linearly graded absorber layers, $\ego$ is at its
minimum allowed value; however, the value of $\egow$ is at its minimum allowed value for the
linearly graded absorber layer (Table~\ref{tab-linear-grading})
but at its maximum allowed value for the homogeneous
absorber layer (Table~\ref{tab-homo-LBC}). The values of $A\sim1$
and $\egmax=1.98$~eV are independent of 
 $\Ls$ for the linearly graded absorber layer (Table~\ref{tab-linear-grading}),
 the latter being significantly lower than its maximum allowed value.

	\begin {table*}[t]
	\caption { \label{tab-linear-grading}
		Predicted parameters of the optimal AlGaAs solar cell with a specified value of $\Ls\in[100,2000]$~nm, 
	when the $n$-AlGaAs  absorber layer is  is linearly nonhomogeneous
		according to Eq.~\r{Eqn:Linear-bandgap1}, $\LAg=100$~nm, and $\zeta<1$. }
	\begin{center}
		\begin{tabular}{ |p{1.1cm}|p{1cm}|p{1.2cm}|p{0.8cm}|p{0.8cm}|p{0.8cm}|p{0.8cm}|p{1.8cm}|p{1.2cm}|p{1cm}|p{1cm}|}
			\hline
			\hline
			$\Ls$ & $\egow$&$\ego$ &$\egmax$ & A & $\Lx$ &$\zeta$  &
			$\Jsc$ & $\Voc$&$FF$&$\eta$  \\
			(nm) & (eV)& (eV)&(eV) &  & (nm) & &(mA~cm$^{-2}$) & (V)& (\%)& (\%) \\
			\hline
		100 & 1.424 &1.424 &1.98&0.99 &500 &0.05 & 16.8 &1.399 & 89.3&21.0\\ \hline
		200 & 1.424 &1.424 &1.98&0.99 &510 &0.05 & 19.0 &1.422 & 81.9&22.2\\ \hline
		300 & 1.424 &1.424 &1.98&1.0  &502 &0.05 & 19.1 &1.441 & 85.3&23.5\\ \hline
		400 & 1.424 &1.424 &1.98&0.99 &510 &0.05 & 19.8 &1.453 & 86.5&24.9\\   \hline 
		500 & 1.424 &1.424 &1.98&0.98 &500 &0.05 & 20.5 &1.462 & 87.1&26.1\\   \hline
		1000 & 1.424 &1.424&1.98&0.99 &500 &0.05 & 22.7 &1.486 & 88.3&29.8\\   \hline
		2000 & 1.424 &1.424&1.98&1.0  &500 &0.05 & 24.7 &1.507 & 88.8&33.1\\  			
			\hline
			\hline
		\end{tabular}
	\end{center}
	\end {table*}

\subsubsection{Detailed study for highest efficiency} \label{ds-linear}
The highest efficiency of $33.1\%$ for the solar cell whose $n$-AlGaAs absorber
layer has a linearly graded bandgap is delivered in Table~\ref{tab-linear-grading}
by the optimal design for 
$\Ls=2000$~nm. We determined the spatial 
profiles of the bandgap $\eg$,   electron affinity  $\chi(z)$,    conduction-band  energy $\Ec(z)$, 
valence-band  energy $\Ev(z)$,    intrinsic energy $\Ei(z)$,
electron density $n(z)$, hole density $p(z)$,  intrinsic charge-carrier density $\ni(z)$,
recombination rate $R(z)$, and  generation rate $G(z)$ in the absorber layer of this
solar cell. Furthermore, we  determined the  total device current density $\Jdev$ delivered to an external 
circuit  as well as the   {electrical} power density   {$P$} as functions of the bias voltage $\Vext$.

\begin{figure}[h]
	\centering   
	\includegraphics[width=0.75\columnwidth]{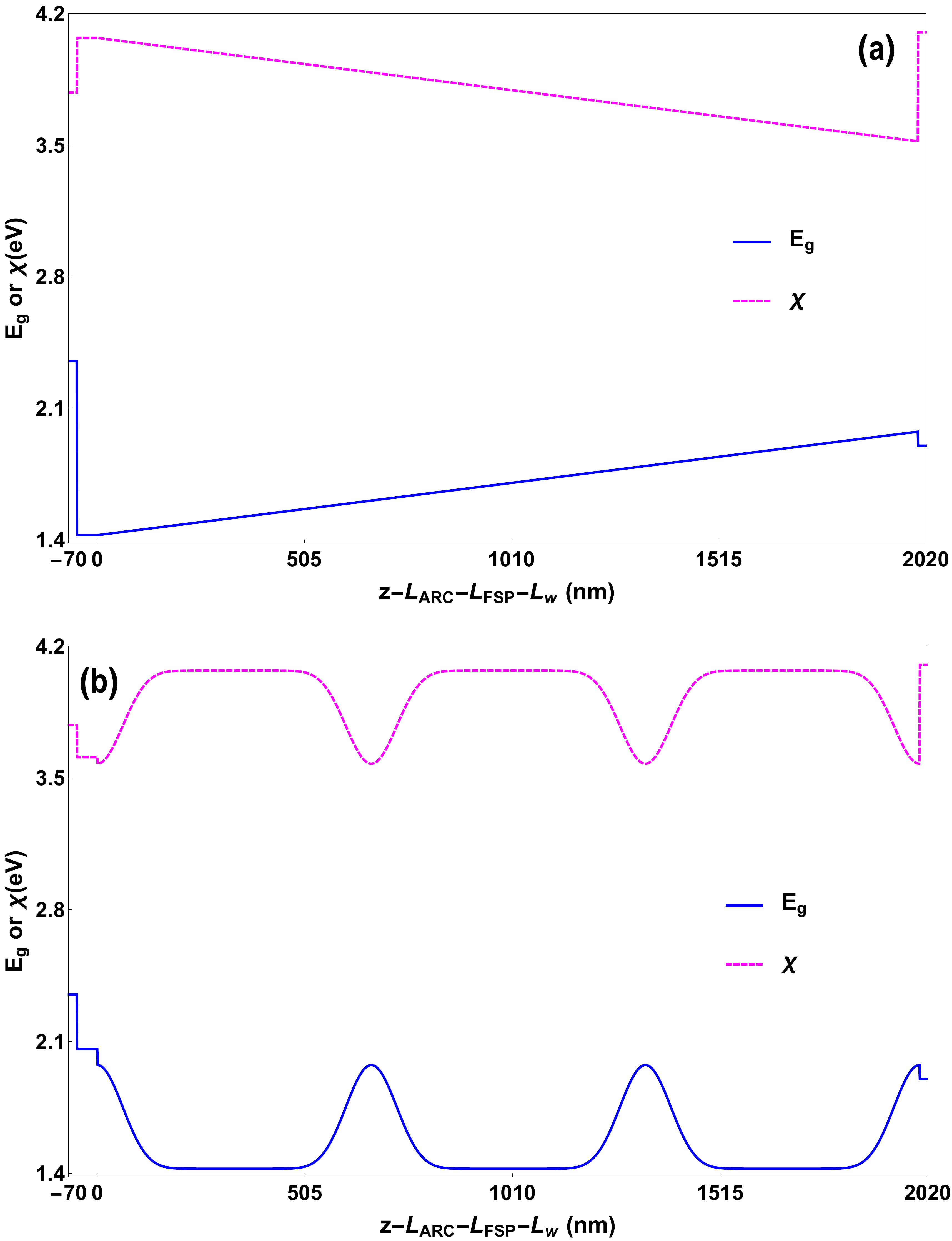} 
	\caption{Spatial profiles of $\eg(z)$ and $\chi(z)$ in the four semiconductor layers of 
		the optimal solar cell with the $2000$-nm-thick $n$-AlGaAs absorber layer with (a) linearly graded   {bandgap and}
		(b) sinusoidally graded bandgap.}
	\label{figure:EgXiSinLinNonH} 
\end{figure}

\begin{figure}[h]
	\centering   
	\includegraphics[width=0.75\columnwidth]{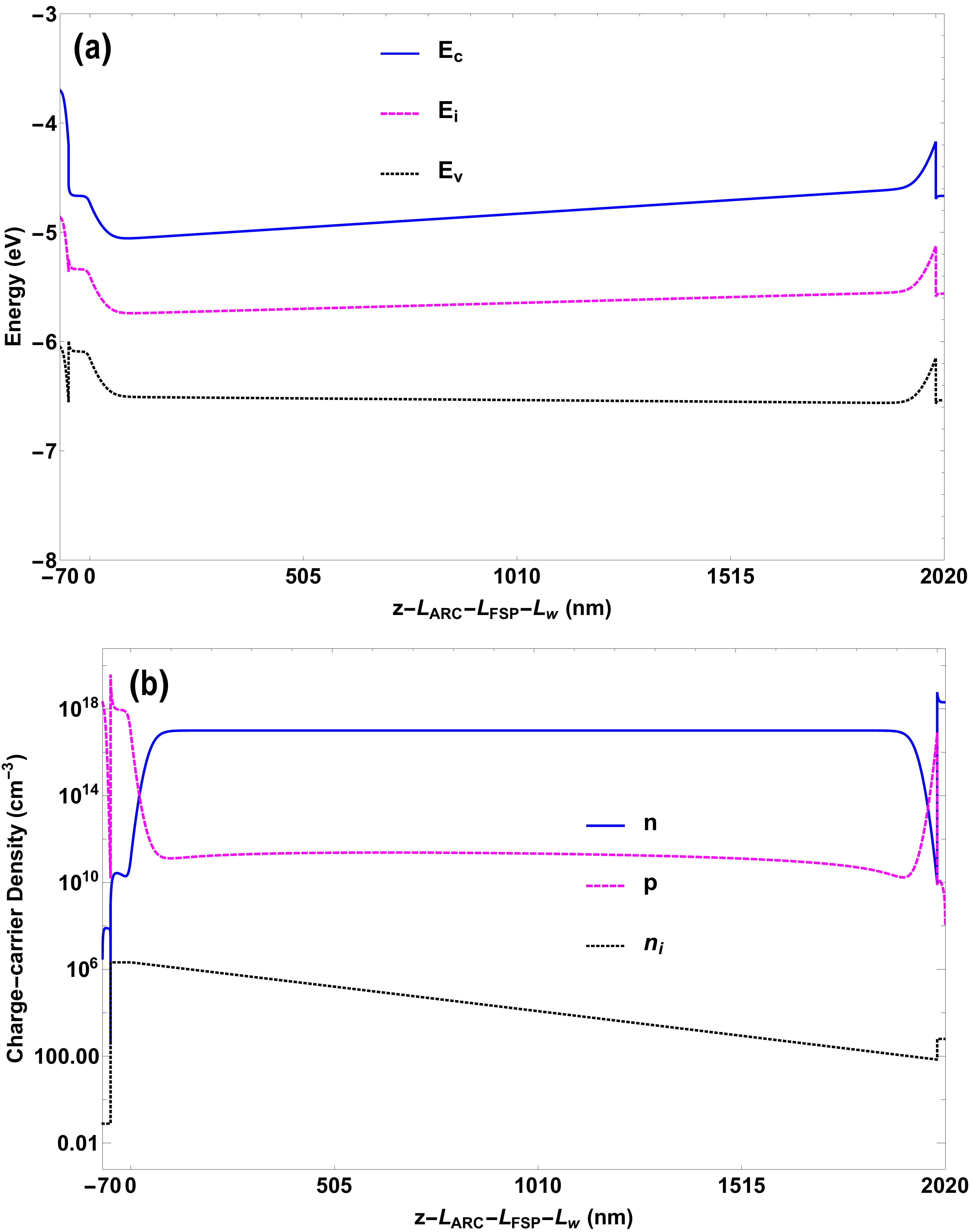} 
	\caption{Spatial profiles of (a) $\Ec(z)$, $\Ev(z)$, and   {$\Ei(z)$, and}
		(b) $n(z)$, $p(z)$, and $\ni(z)$ in the four semiconductor layers of 
		the optimal solar cell with the $2000$-nm-thick $n$-AlGaAs absorber layer with linearly graded bandgap. }
	\label{figure:EcEvEinpnilinNonH} 
\end{figure}

Spatial profiles of $\eg(z)$ and $\chi(z)$  for the optimal solar cell with $n$-AlGaAs absorber
layer of thickness $\Ls=2000$~nm are provided in   {Fig.~\ref{figure:EgXiSinLinNonH}(a),
whereas} Fig.~\ref{figure:EcEvEinpnilinNonH}(a) presents the spatial profiles of  
$\Ec(z)$, $\Ev(z)$, and $\Ei(z)$.  The spatial variations of $\Ec$ and $\Ei$ are similar to that
of $\eg$   {[Fig.~\ref{figure:EgXiSinLinNonH}(a)]}.
Figure~\ref{figure:EcEvEinpnilinNonH}(b) presents the spatial profiles of $n(z)$, $p(z)$, and $\ni(z)$
in steady-state condition. The intrinsic carrier density varies linearly such that it is small
where $\eg$ is large and \textit{vice versa}. 

\begin{figure}[h]
	\centering   
	\includegraphics[width=0.75\columnwidth]{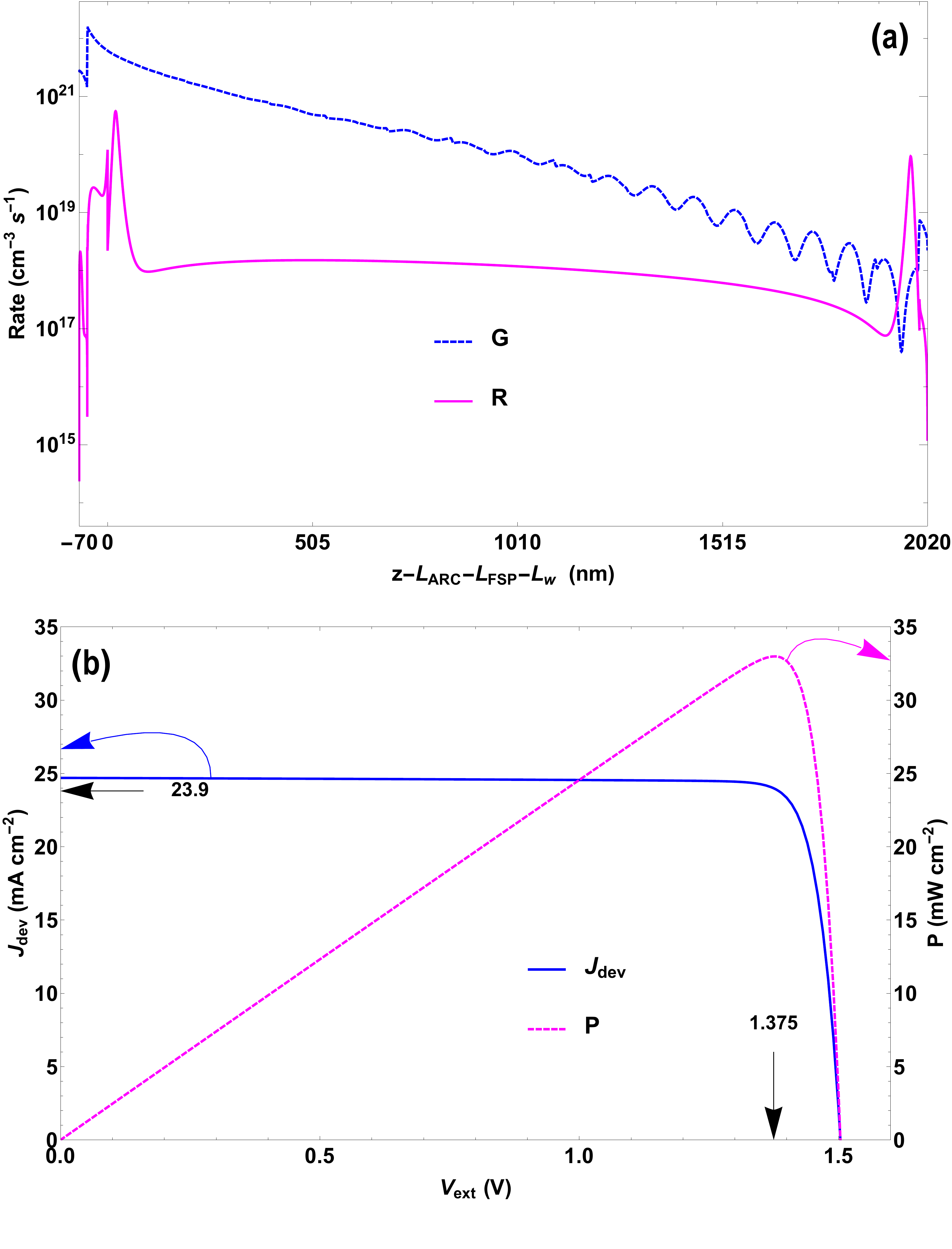} 
	\caption{(a) Spatial profiles of $G(z)$ and $R(z)$ in the four semiconductor layers of 
		the optimal solar cell with the $2000$-nm-thick $n$-AlGaAs absorber layer with linearly graded bandgap. 
		(b)   $\Jdev$-$\Vext$ and $P$-$\Vext$ curves of this solar cell. The numerical values 
		of $\Jdev$ and $\Vext$ for maximum $P$ are also identified. }
	\label{GRJV-linNonH} 
\end{figure}

Spatial profiles of 
$G(z)$ and $R(z)$ are given in Fig.~\ref{GRJV-linNonH}(a).
The  generation rate is higher near the front face and lower near the back face of the $n$-AlGaAs absorber layer, 
which is in accord~\cite{Fonash} with higher   {electron--hole-pair} generation where $\eg$ is lower and \textit{vice versa}. 
Finally, the $\Jdev$-$\Vext$ characteristics of the solar cell   shown in Fig.~\ref{GRJV-linNonH}(b) deliver
$\Jdev=23.9$~mA~cm$^{-2}$ and $\Vext=1.375$~V  for best performance (i.e., for maximum $P$).	

\subsection{Optimal AlGaAs solar cell: Sinusoidally graded bandgap}\label{Sec:Sinusoidally-graded}

\subsubsection{Optimal designs}\label{od-sinusoidal}
Finally, we considered the maximization of $\eta$ when the bandgap of the $n$-AlGaAs 
absorber layer is sinusoidally  graded according to Eq.~(\ref{Eqn:Sin-bandgap}), $\LAg=100$~nm,
and $\zeta\ne 1$. 
The parameter space used for optimizing $\eta$ was chosen as:  
$\egow\in[1.424, 2.09]$~eV, $\ego\in[1.424, 2.09]$~eV, $\egmax\in[1.424, 2.09]$~eV, $A\in[0,1]$,  
$\alpha\in[0, 8]$, $K\in[0, 8]$,  $\psi\in[0,1]$,   {$\Lx\in[100, 1000]$~nm}, and $\zeta\in[0.05,1]$.  
 
Values of  $\Jsc$, $\Voc$, $FF$, and $\eta$ predicted by our 
model presented in Table~\ref{tab-sinu-period} for seven different values of $\Ls$.
The   values of $\egow$, $\ego$, $\egmax$, $A$, $\alpha$, $K$, $\psi$, $\Lx$, and 
$\zeta$ for the optimal designs are also provided in the same table.

For the thinnest $n$-AlGaAs absorber layer ($\Ls=100$~nm), the maximum efficiency predicted is $21.2\%$ 
with $\egow=2.09$~eV ($\xib=0.8$), $\ego=1.424$~eV ($\xi=0$), $\egmax=1.98$~eV ($\xi=0.45$),   {$A=1$},  $\alpha=6$, $K=3$, 
$\psi=0.75$, and $\Lx=510$~nm. 
The values of $\Jsc$, $\Voc$, and $FF$ corresponding to this optimal design are   {$16.1$~mA~cm$^{-2}$}, $1.455$~V, 
and $90.3\%$, respectively. A relative enhancement of $14.5$\% over the optimal efficiency $18.5\%$ for the
homogeneous $n$-AlGaAs absorber layer (Table~\ref{tab-homo-LBC})   is predicted.

For the thickest $n$-AlGaAs absorber layer ($\Ls=2000$~nm), the maximum efficiency predicted is $34.5\%$ 
with $\egow=2.09$~eV ($\xib=0.8$), $\ego=1.424$~eV ($\xi=0$), $\egmax=1.98$~eV ($\xi=0.45$), $A=0.99$, 
$\alpha=6$, $K=3$, $\psi=0.75$, and $\Lx=550$~nm.  
The corresponding values of $\Jsc$, $\Voc$, and $FF$   are   {$24.8$~mA~cm$^{-2}$}, $1.556$~V, 
and $89.2\%$, respectively.
A relative enhancement of $19.8\%$ is predicted with sinusoidal grading of $n$-AlGaAs absorber layer over the 
optimal efficiency of $28.8\%$ with homogeneous $n$-AlGaAs absorber layer (Table~\ref{tab-homo-LBC}).
Just as in Sec.~3.\ref{od-linear},
although $\Voc$ is significantly higher with the sinusoidally graded bandgap   compared to the homogeneous bandgap  (Table~\ref{tab-homo-LBC}),   $\Jsc$ is
lower with the sinusoidally graded bandgap.

The optimal designs in Table~\ref{tab-sinu-period} have $\Lx=525\pm25$~nm and $\zeta=0.05$. 
The values of $\ego$, $\egmax$, $A$, $\alpha$, and   {$\psi$} are the same for all values of $\Ls$; 
however, $K\in\left\{1,2,3\right\}$ does vary with  $\Ls$. 
The values of $A\sim1$
and $\egmax=1.98$~eV are independent of 
 $\Ls$ for the sinusoidally graded absorber layer (Table~\ref{tab-sinu-period}),
 the latter being significantly lower than its maximum allowed value.

The highest possible efficiency ($34.5\%$) with a sinusoidally graded $n$-AlGaAs absorber layer
is $4.2\%$ higher than the highest possible efficiency ($33.1\%$) with a linearly graded
$n$-AlGaAs absorber layer (Table~\ref{tab-linear-grading}).
The short-circuit current density is almost the same
for both   linearly and sinusoidal graded $n$-AlGaAs 
absorber layers; however, the open-circuit voltage is somewhat higher for sinusoidally graded $n$-AlGaAs absorber layer.  
By comparing Tables~\ref{tab-linear-grading} and \ref{tab-sinu-period},  
we   conclude that sinusoidally graded $n$-AlGaAs absorber layer leads
to significantly higher efficiency than the   
 linearly graded $n$-AlGaAs absorber layer for $\Ls\geq1000$~nm,
but both types of graded-bandgap absorber layers deliver practically the same efficiency
for $\Ls\leq500$~nm.

	\begin {table*}[t]
	\caption {\label{tab-sinu-period} 
	Predicted parameters of the optimal AlGaAs solar cell with a specified value of $\Ls\in[100,2000]$~nm, 
	when the $n$-AlGaAs  absorber layer is sinusoidally graded
		according to Eq.~\r{Eqn:Sin-bandgap}, $\LAg=100$~nm, and $\zeta<1$. }
	\begin{center}
		\begin{tabular}{ |p{1.1cm}|p{0.7cm}| p{0.7cm} | p{0.7cm}|p{0.8cm}| p{0.8cm}| p{0.8cm}|p{0.8cm}|p{0.8cm}|p{0.8cm}|p{1cm}|p{1cm}|p{1cm}|p{1cm}|}
			\hline
			\hline
			$\Ls$ & $\egow$& $\ego$ & $\egmax$ & A &$\alpha$& $K$ & $\psi$ & $\Lx$ &$\zeta$  &
			$\Jsc$ & $\Voc$&  $FF$ &$\eta$  \\
			(nm) & (eV) &(eV)& (eV)& & & & & & &(mA&   {(V)}&(\%)&(\%)\\
			&   & & & & & & &    {(nm)}  & & cm$^{-2}$)& & &  \\
			\hline
			100  & 2.09 &1.424 &1.98 &1.0 & 6 & 3   & 0.75 &510 &0.05  & 16.1  &1.455  &90.3 &21.2\\ \hline
			200  & 2.09 &1.424 &1.98 &1.0 & 6 & 1   & 0.75 &520 &0.05  & 19.2  &1.471  &80.2 &22.6 \\	\hline			
			300  & 2.06 &1.424 &1.98 &1.0 & 6 & 1   & 0.74 &512 &0.05  & 19.7  &1.486  &80.2 &23.5\\ \hline
			400  & 2.09 &1.424 &1.98 &1.0 & 6 & 1   & 0.75 &509 &0.05  & 20.2  &1.497  &82.0 &24.8\\  \hline
			500  & 2.08 &1.424 &1.98 &1.0 & 6 & 1   & 0.75 &524 &0.05  & 20.8  &1.505  &83.0 &26.0\\ \hline
			1000 & 2.09 &1.424 &1.98 &1.0 & 6 & 2   & 0.75 &516 &0.05  & 22.5  &1.533  &87.8 &30.4\\ \hline	
			2000 & 2.09 &1.424 &1.98&0.99 & 6 & 3   & 0.75 &550 &0.05  & 24.8  &1.556  &89.2 &34.5\\  
			\hline
			\hline
		\end{tabular}
	\end{center}
	\end {table*}

\subsubsection{Detailed study for highest efficiency}\label{ds-sinusoidal}
We performed a detailed study for the  solar cell with 
thickest ($\Ls=2000$~nm) sinusoidally graded $n$-AlGaAs absorber layer,
because it delivers the highest efficiency.
The variations of $\eg$ and $\chi$  with $z$ in 
the  semiconductor region are provided 
in Fig.~\ref{figure:EgXiSinLinNonH}(b).
The magnitude of $\eg$ is large near both faces of the $n$-AlGaAs absorber layer, 
which features elevate $\Voc$~\cite{Ahmad2019}. The regions
in which $\eg$ is small are of substantial thickness, these regions being
responsible for elevating $G(z)$~\cite{Fonash}.

\begin{figure}[h]
	\centering   
	\includegraphics[width=0.75\columnwidth]{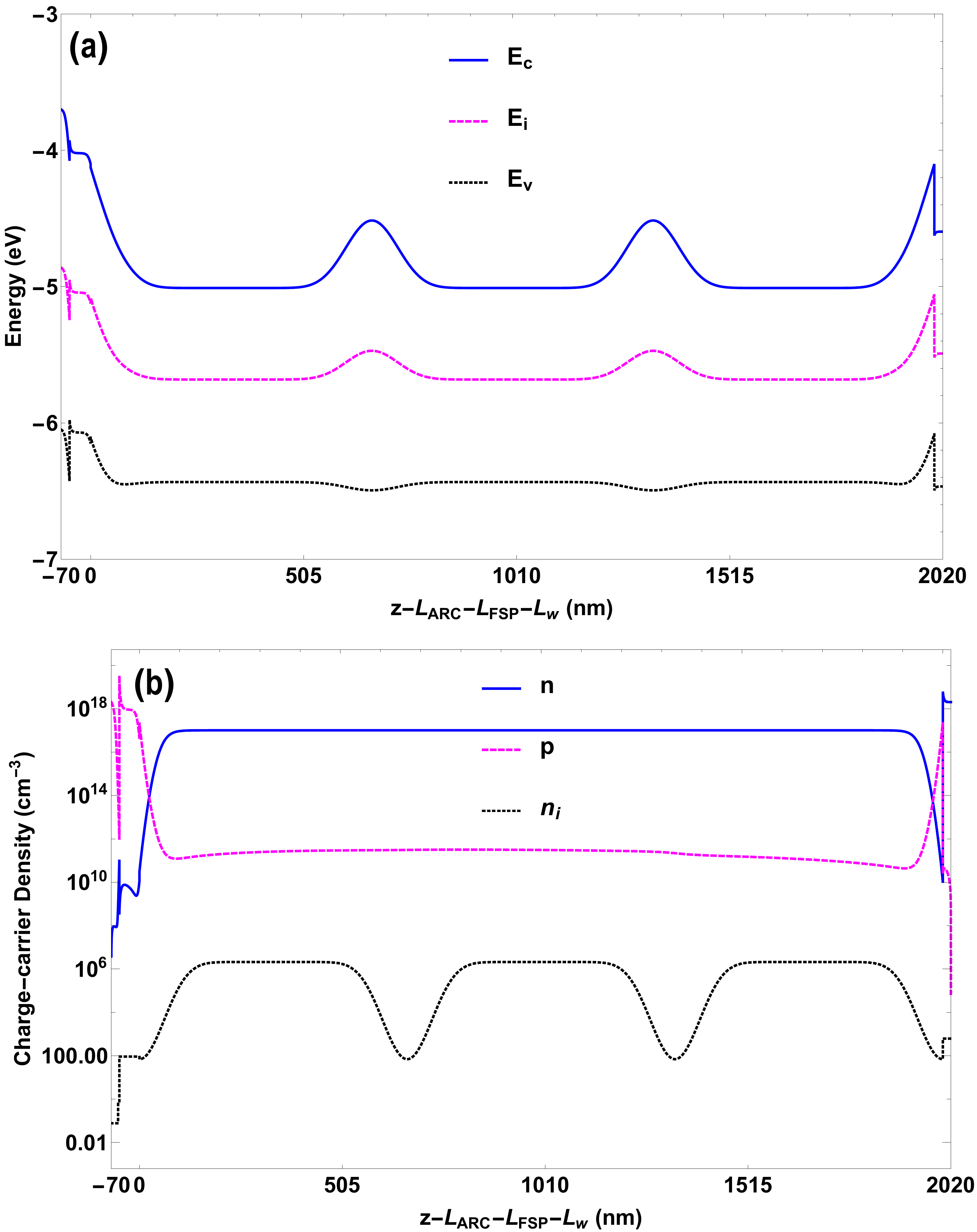} 
	\caption{Spatial profiles of (a) $\Ec(z)$, $\Ev(z)$, and   {$\Ei(z)$, and}
		(b) $n(z)$, $p(z)$, and $\ni(z)$ in the four semiconductor layers of 
		the optimal solar cell with the $2000$-nm-thick $n$-AlGaAs absorber 
		layer with sinusoidally graded bandgap. }
	\label{figure:EcEvEinpniSinNonH} 
\end{figure}

\begin{figure}[h]
	\centering   
	\includegraphics[width=0.75\columnwidth]{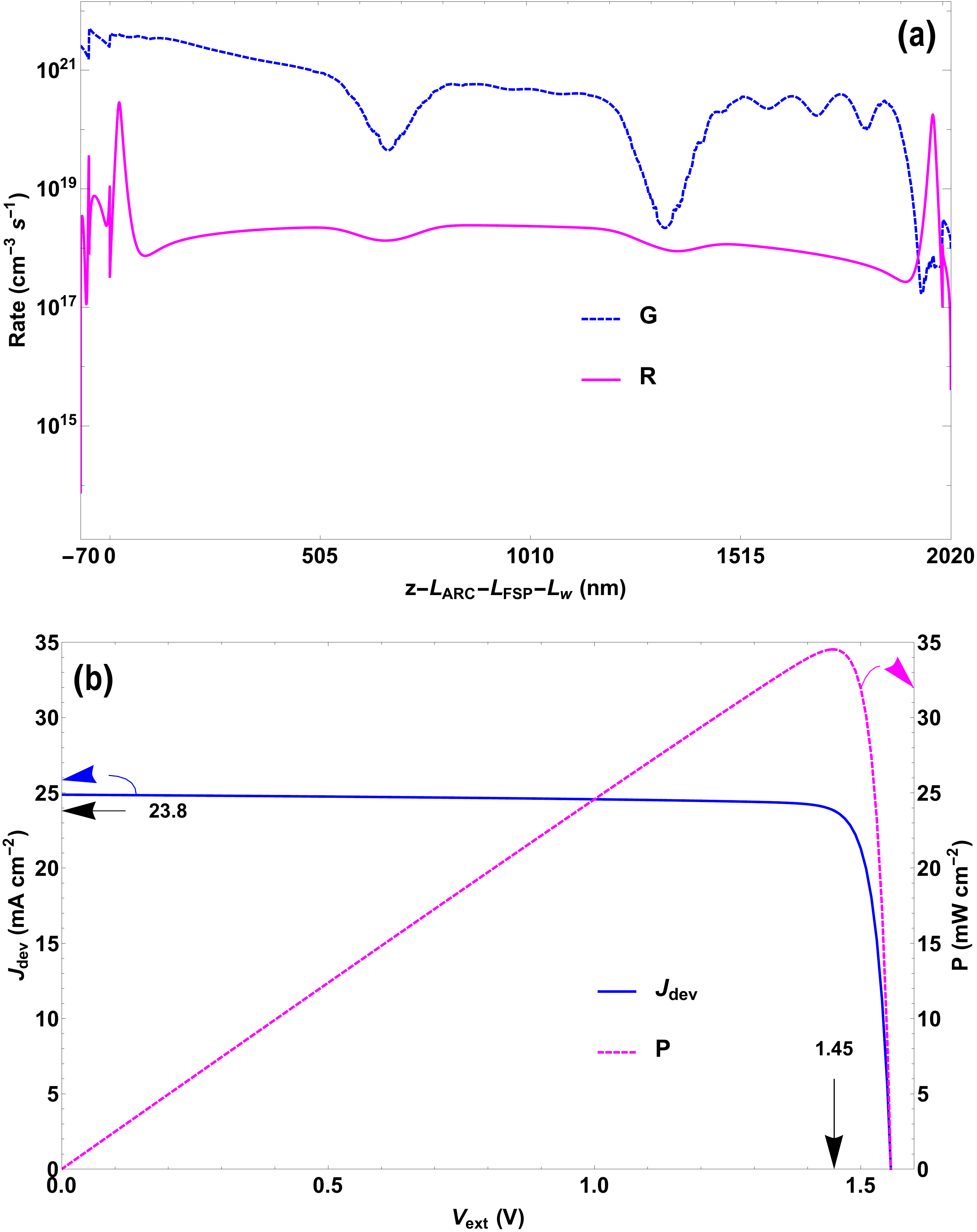} 
	\caption{(a) Spatial profiles of  $G(z)$ and $R(z)$ in the four semiconductor layers of 
		the optimal solar cell with the $2000$-nm-thick $n$-AlGaAs absorber layer with sinusoidally graded bandgap. 
		(b)   $\Jdev$-$\Vext$ and $P$-$\Vext$ curves of this solar cell. The numerical values 
		of $\Jdev$ and $\Vext$ for maximum $P$ are also identified. }
	\label{figure:GRJV-SinNonH} 
\end{figure}

Figure~\ref{figure:EcEvEinpniSinNonH}(a) shows 
the variations of $\Ec$, $\Ev$, and $\Ei$ with respect to $z$.
The spatial profiles of $\Ec$ and $\Ei$ are similar to that of $\eg$.
Figure~\ref{figure:EcEvEinpniSinNonH}(b) shows the spatial variations
of the electron, hole, and intrinsic carrier densities in steady-state condition.  
The intrinsic carrier density  varies sinusoidally such that
$\ni$ is small  where $\eg$ is large and \textit{vice versa}.    
Profiles of $G(z)$ and 
$R(z)$ are shown in Fig.~\ref{figure:GRJV-SinNonH}(a). 
The generation rate is higher in regions with
lower bandgap and \textit{vice versa}.
The $\Jdev$-$\Vext$ characteristics of the solar cell are shown in 
Fig.~\ref{figure:GRJV-SinNonH}(b).  Our optoelectronic
model predicts $\Jdev=23.8$~mA~cm$^{-2}$ and $\Vext=1.45$~V  for best performance.

\section{Concluding remarks}\label{sec:conc}

A coupled optoelectronic model along with the differential evolution algorithm was implemented to 
evaluate the effectiveness of grading the bandgap of the $n$-AlGaAs absorber layer
for improving the power conversion efficiency of thin-film AlGaAs solar cells. 
Both linearly and sinusoidally graded bandgaps were studied, with the semiconductor region
of the solar cell backed  by a periodically corrugated Ag backreflector combined with 
the localized ohmic Pd--Ge--Au backcontacts.

A $2000$-nm-thick $n$-AlGaAs absorber layer that is sinusoidally 
graded can deliver
$34.5$\% efficiency, $24.8$~mA~cm$^{-2}$ short-circuit current density, $1.556$~V open-circuit voltage, 
and  $89.2$\% fill factor. 
In comparison, the  efficiency is $28.8$\%, the short-circuit current density is  $30.2$~mA~cm$^{-2}$,  
the open-circuit voltage is $1.090$~V, and the fill factor
is $87.3$\% when the bandgap of the absorber layer is homogeneous. 
Efficiency enhancement can also be achieved by linearly grading the bandgap of the $n$-AlGaAs absorber layer, 
but the gain is significantly smaller compared to sinusoidal bandgap grading when the absorber
layer is at least 1000-nm thick.
However, for thinner $n$-AlGaAs absorber layers,
both linearly graded bandgaps and sinusoidally   {graded bandgaps can} provide almost equal efficiency gains over the homogenous bandgap.

When the bandgap is sinusoidally graded in the $n$-AlGaAs absorber layer, the   {electron--hole-pair} generation rate is 
higher in the broad small-bandgap regions than elsewhere in the $n$-AlGaAs absorber layer~\cite{Fonash}.  
The open-circuit voltage is elevated in the optimal designs~\cite{Ahmad2019}, because the bandgap is high 
in the vicinity of both faces of the $n$-AlGaAs absorber layer. Both of these characteristics help to increase the  
 efficiency.

Optoelectronic optimization thus indicates that $34.5$\% efficiency (Table~\ref{tab-sinu-period}) can be achieved for AlGaAs solar cell 
with a $2000$-nm-thick sinusoidally graded $n$-AlGaAs absorber layer. This efficiency 
  {is significantly} higher compared 
to $27.4$\% efficiency demonstrated with the homogeneous $n$-AlGaAs absorber layer with a
continuous ohmic Pd--Ge--Au back contact (Table~\ref{tab--ref-results}).
Efficiency improvements of
equivalent magnitude---e.g., from $22\%$ to $27.7\%$---have been predicted by bandgap grading of the CIGS absorber layer in 
thin-film CIGS solar cells~\cite{Ahmad2019}. Thus, bandgap grading can provide a way to realize more efficient 
thin-film solar cells for  ubiquitous harnessing of solar energy at  low-wattage levels. 

\appendix 
\section{Optical and electrical parameters}\label{Appendix A}

\begin{figure}[h]
	\centering
	\includegraphics[width=\columnwidth]{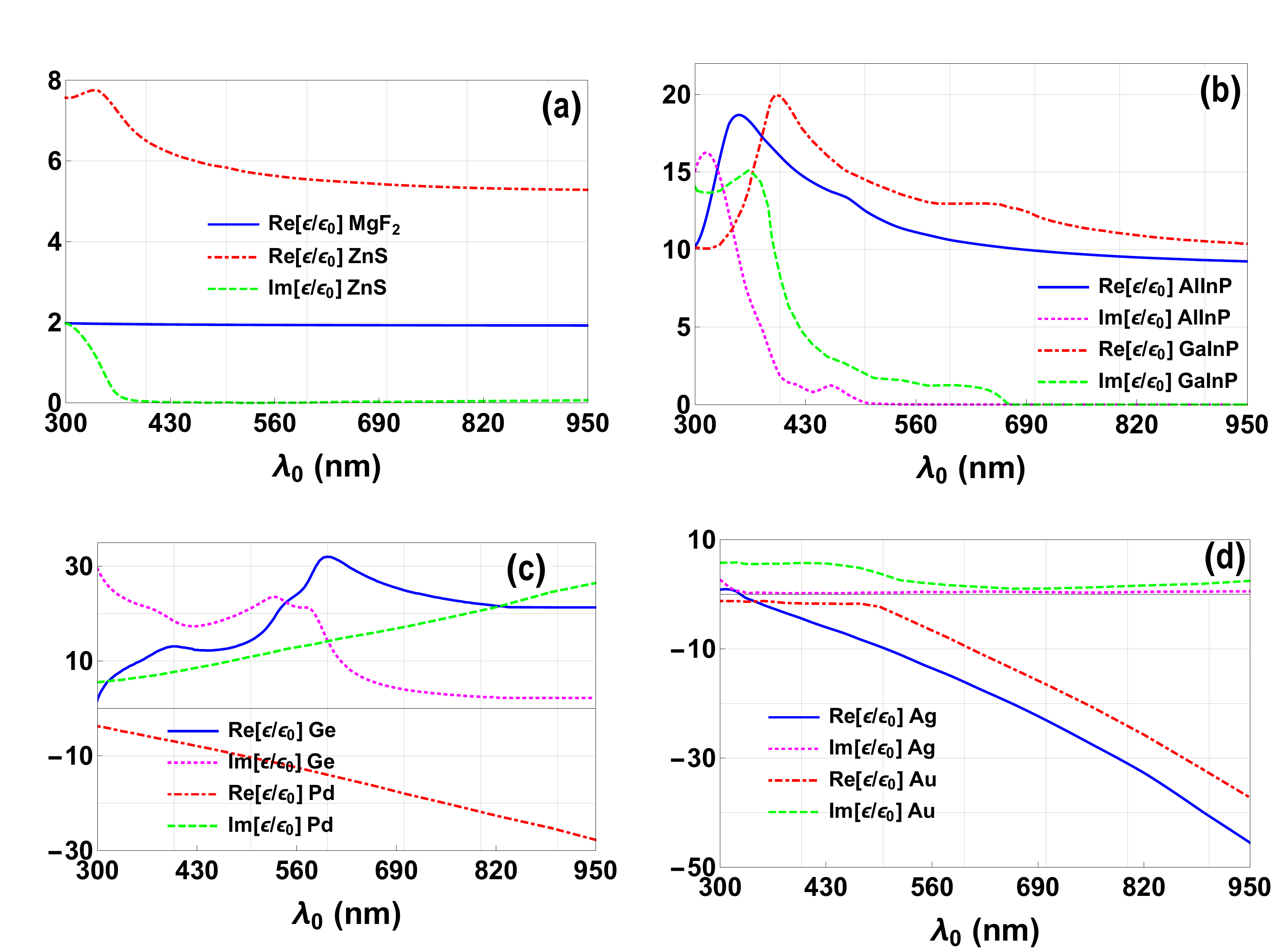}    
	\caption{Real and  imaginary parts of the optical relative permittivity $\eps/\epso$ of (a)
		MgF$_2$, ZnS, (b) AlInP, GaInP, (c) Pd, Ge,  (d) Au, and Ag
		as functions of $\lambdao\in[300,950]$~nm, with $\epso$ denoting
		the permittivity of free space. The
		imaginary part of the relative permittivity of MgF$_2$ is negligibly
		small. 
		\label{spectra-eps-1}}
\end{figure}

\begin{figure}[h]
	\centering
	\includegraphics[width=\columnwidth]{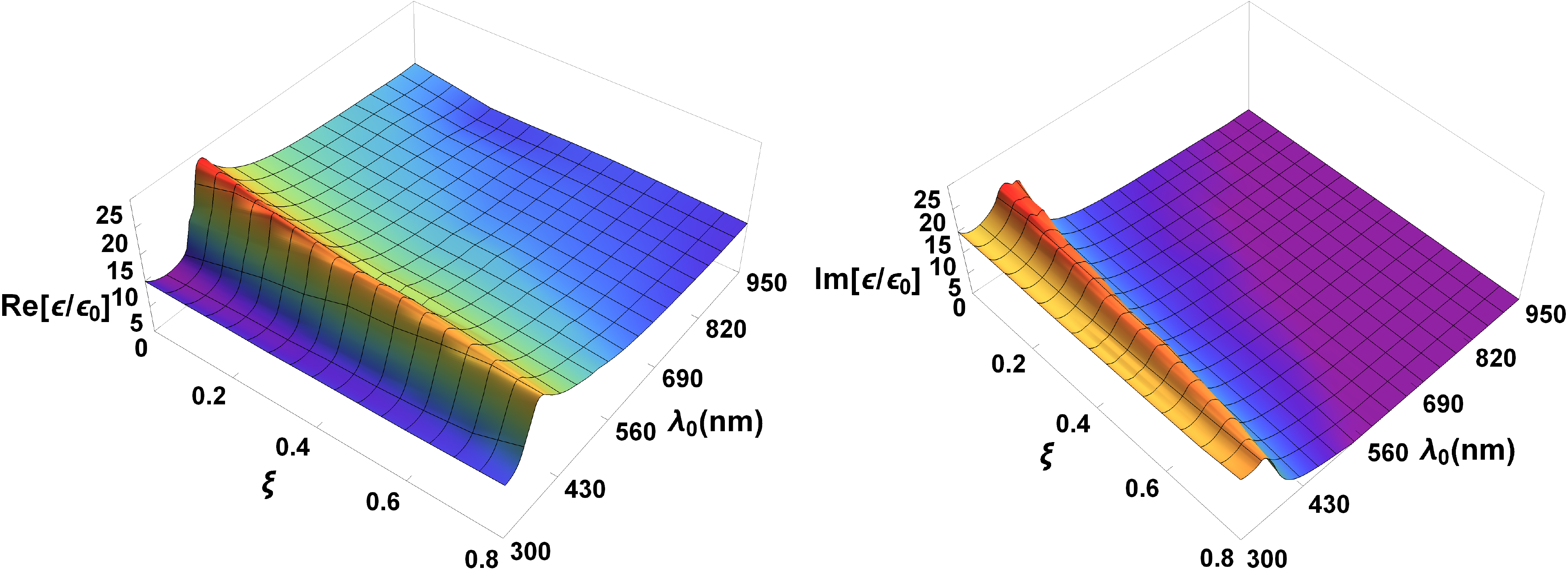}    
	\caption{Real and  imaginary parts of the optical relative permittivity $\eps/\epso$ of 
		AlGaAs as functions of $\lambdao\in[300,950]$~nm and  $\xi\in[0,0.8]$. 
		\label{AlGaAs-spectra}}
\end{figure}

The optical permittivity $\eps$ of any material is, in general, a function
of $\lambdao$.
The  optical relative permittivities of 
MgF$_2$~\cite{mgf2}, ZnS~\cite{ZnS}, AlInP~\cite{AlInP}, GaInP~\cite{GaInP}, Pd~\cite{JohnsonChristy}, Ge~\cite{Jellison1992}, Au~\cite{JohnsonChristy}, and Ag~\cite{JohnsonChristy}
are provided in Fig.~\ref{spectra-eps-1} for $\lambdao\in[300,950]$~nm. The
real and  imaginary parts of the optical relative permittivity  of AlGaAs
are provided in Fig.~\ref{AlGaAs-spectra}
as functions of $\lambdao\in[300,950]$~nm and $\xi\in[0,0.8]$~\cite{Aspnes1986},
data being  unavailable for $\xi\in(0.8,1]$.

Table~\ref{tab-elec-prop} provides the values of electrical parameters used for all four semiconductors   {\cite{Vurgaftman2001, Gudovskikh2007, Adachi1985, Adachi-book, ioffe}}.

\begin{widetext}
	\begin{table}[h] 
		\caption{Electrical properties of AlInP~  {\cite{Vurgaftman2001, Gudovskikh2007}}, GaInP~  {\cite{ioffe,Vurgaftman2001, Gudovskikh2007}} and Al$_{\xi}$Ga$_{1-\xi}$As~  {\cite{Adachi1985,Adachi-book, ioffe}}.
			\label{tab-elec-prop}}
		\centering
		\begin{tabular}{|p{2.6cm}|p{2.2cm}|p{2.2cm}|p{2.8cm}|p{6.2cm}|}\hline
			Parameter  & Symbol~(unit) & AlInP~  {\cite{Vurgaftman2001, Gudovskikh2007}}&GaInP~  {\cite{ioffe,Vurgaftman2001, Gudovskikh2007}} &Al$_{\xi}$Ga$_{1-\xi}$As~  {\cite{Adachi1985,Adachi-book, ioffe}}\\\hline\hline
			Bandgap  & $\eg$~(eV) &$2.35$&$1.9$ & $1.424 + 1.247\xi$, $0\leq\xi<0.45$;\\
			&&&&$1.9+ 0.125\xi + 0.143\xi^2$, $0.45\leq\xi\leq 1$\\
			Electron affinity & $\chi$~(eV)&$3.78$ &$4.1$ & $4.07 - 1.1\xi$, 0$\leq\xi<0.45$;\\ 
			&&&&$3.64 - 0.14\xi$, $0.45\leq\xi\leq1$\\
			Doping density  & $\ND$~(cm$^{-3}$)&$2\times10^{18}$ (acceptor)&$2\times10^{18}$ (donor) &1$\times10^{18}$ (acceptor and donor)\\
			
			Conduction-band & $\Nc$~(cm$^{-3}$)&$2.5\times 10^{18}$&$6.5\times 10^{17}$&$2.5\times10^{19}(0.063+0.083\xi)^{3/2}$, $0\leq\xi<0.45$;\\
			density of states&&&&$2.5\times 10^{19} (0.85-0.14\xi$)$^{3/2}$, $0.45\leq\xi<1$\\
			
			Valence-band density of states & $\Nv$~(cm$^{-3}$) & $7\times 10^{18}$&$1.5\times 10^{19}$ &$2.5\times10^{19}(0.51+0.25\xi)^{3/2}$\\

			Electron mobility & $\mun$~(cm$^2$V$^{-1}$s$^{-1}$)&$100$ &$500$ &$8\times 10^3-2.2\times 10^4 \xi+10^4\xi^2$, $0\leq\xi<0.45$;\\	
			&&&&$-255+1160\xi-720\xi^2$, $0.45\leq\xi\leq1$\\
			Hole mobility & $\mup$~(cm$^2$V$^{-1}$s$^{-1}$)&$10$ &$30$ &$370-970\xi+740\xi^2$\\
			DC relative permittivity  & $\epsdc$&$11.8$& $11.8$&$ 13.18-3.12\xi$\\
			Defect/trap density &$\NT$~(cm$^{-3}$)&$10^{17}$ &$10^{17}$ & $(1+9\xi)\times10^{15}$\\
			Defect/trap level &$\ET$~(eV)& Midgap&Midgap & $0.75$~eV below conduction-band energy \\
			Electron capture cross section& $\sigman$~(cm$^2)$&$10^{-14}$& $10^{-14}$& $10^{-16}$\\
			Hole capture cross section & $\sigmap$~(cm$^2)$&$10^{-14}$ &$10^{-14}$& $10^{-16}$\\
			Radiative recombination coefficient & $\RB$~(  {cm$^{3}$}~s$^{-1}$) &$10^{-10}$ &$10^{-10}$&$1.8\times10^{-10}$\\
			Electron thermal speed & $\vthn$~(cm~s$^{-1}$) &$10^{7}$ &$10^{7}$&$(4.4-2.1\xi)\times10^{7}$\\
			Hole thermal speed & $\vthp$~(cm~s$^{-1}$) &$10^{7}$ &$10^{7}$&$(1.8-0.5\xi)\times10^{7}$\\
			Auger electron recombination coefficient & $C_{\rm n}$~(cm$^{6}$~s$^{-1}$) &$10^{-30}$ &$10^{-30}$&$10^{-30}$\\
			Auger hole recombination coefficient & $C_{\rm p}$~(cm$^{6}$~s$^{-1}$) &$10^{-30}$ &$10^{-30}$&$10^{-30}$\\
			\hline\hline
		\end{tabular}
	\end{table}
\end{widetext}

\vspace{3mm}	

\noindent\textbf{Funding.} 
The research of  F. Ahmed 
and A. Lakhtakia was partially supported by  US National Science Foundation (NSF)
under grant number DMS-1619901. 
The research of P.B.  Monk was partially supported by  the US
National Science Foundation (NSF) under grant number DMS-1619904. 

\noindent\textbf{Acknowledgments.}
A. Lakhtakia thanks the Charles Godfrey Binder Endowment at the Pennsylvania State University  and  the 
Otto M{\o}nsted Foundation for partial support of his research endeavors.

\noindent\textbf{Disclosures.} The authors declare no conflicts of interest.

\end{document}